\documentclass[10pt, a4paper, conference, compsocconf]{IEEEtran}
\usepackage{amsmath}
\usepackage{graphicx}
\usepackage{amsfonts}
\usepackage{amssymb}
\usepackage{alltt}
\usepackage{stmaryrd}
\usepackage{dsfont} 
\usepackage{color} 
\usepackage[standard]{ntheorem}
\usepackage{floatflt}
\usepackage{caption}
\usepackage[font=footnotesize]{subfig}
\usepackage{array}
\usepackage{multirow,booktabs}
\renewcommand{\theequation}{\arabic{equation}}
\usepackage[width=18.2cm,height=25.7cm]{geometry}

\ifCLASSINFOpdf
\else
\fi
\begin{document}

\title{ Improving random number generators by chaotic iterations \\ Application in data hiding}
\author{\IEEEauthorblockN{Christophe Guyeux\IEEEauthorrefmark{1},
Qianxue Wang\IEEEauthorrefmark{1} and
Jacques M. Bahi\IEEEauthorrefmark{1}}
\IEEEauthorblockA{\IEEEauthorrefmark{1}University of Franche-Comte\\
Computer Science Laboratory LIFC,
Belfort, France\\ Email: christophe.guyeux@univ-fcomte.fr, qianxue.wang@univ-fcomte.fr, jacques.bahi@univ-fcomte.fr}}

\maketitle

\begin{abstract}
In this paper, a new pseudo-random number generator\\
(PRNG) based on chaotic iterations is proposed. This method also combines the digits of two XORshifts PRNGs. The statistical properties of this new generator are improved: the generated sequences can pass all the DieHARD statistical test suite. In addition, this generator behaves chaotically, as defined by Devaney. This makes our generator suitable for cryptographic applications. An illustration in the field of data hiding is presented and the robustness of the obtained data hiding algorithm against attacks is evaluated.


\end{abstract}

\begin{IEEEkeywords}
Chaotic sequences; Topological chaos; Pseudo-random number generator; Statistical tests; Internet security; data hiding; Discrete chaotic iterations.

\end{IEEEkeywords}

\IEEEpeerreviewmaketitle

\section{Introduction}
A pseudo-random number generator (PRNG) is an algorithm for generating a sequence of numbers that is supposed to be indistinguishable from a uniformly chosen random sequence~\cite{L'ecuyer2008}. The sequence is not truly random in that it is completely determined by a relatively small set of initial values, called the PRNG's seed. Compare to hardware-based approaches, these PRNGs must be easy to generate and process, but are less closer to truly random behavior. PRNGs play an important role in practice for a whole range of applications such as information security, statistics (samplings, simulations and Monte-Carlo techniques), video games and gambling machines, to name a few~\cite{Gonzaleza2005}. These PRNGs are often based on logical operations like bitwise exclusive or (XOR) and on circular shift of bit vectors. XORshift, designed by George Marsaglia~\cite{Marsaglia2003}, is a popular example of such generators. However, the security level of some PRNGs of this kind has been revealed to be inadequate by today's standards. We investigated whether it would be possible to combine two generators in some way that would give better properties than the individual components alone.\newline
This paper extends the study started in \cite{guyeux09} and~\cite{wang2009}. In \cite{guyeux09}, it is proven that chaotic iterations (CIs), a suitable tool for fast computing iterative algorithms, satisfies the chaos property, as it is defined by Devaney~\cite{Dev89}. In \cite{wang2009}, the chaotic behavior of CIs are exploited in order to obtain an unpredictable PRNG. This generator, based on chaotic iterations, depends on two input sequences. In \cite{wang2009}, these two sequences are constituted by two logistic maps. This novel generator has successfully passed the NIST (National Institute of Standards and Technology of the U.S. Government) battery of tests~\cite{ANDREW2008}.\newline
In this new paper, we achieve to improve the speed of the former PRNG, by using two XORshifts in place of the logistic map. In addition, this new version of our PRNG is able to pass the famous DieHARD statistical battery of tests~\cite{Marsaglia1996}.
And its security is improved compared to XORshift alone, and to our former PRNG. After presenting the theoretical framework of the study, 
 a concrete example of how to use these pseudo-random numbers in the field of data hiding is detailed. An analysis focuses on the watermarked images which have already been subjected to common image distortion attacks. It is shown that sequences generated from this generator have a good robustness in the presence of such attacks.\newline
The rest of this paper is organized in the following way: in Section~\ref{Basic recalls}, some basic definitions concerning chaotic iterations and PRNGs are recalled. Then, the generator based on discrete chaotic iterations is presented in Section~\ref{The generation of pseudo-random sequence}. In Section~\ref{Testing a generator}, we show that the proposed PRNG passes the DieHARD statistical tests. In Sections~\ref{An application example of the proposed PRNG} and~\ref{Robustness evaluations}, there is a discussion on a potential application scenario to watermarking. Finally, some conclusions and future work are drawn in Section~\ref{Conclusions and Future Work}.
\section{Basic recalls}
\label{Basic recalls}
This section is devoted to basic notations and terminologies in the fields of chaotic iterations and PRNGs.
\subsection{Notations}
\begin{tabular}{@{}c@{}@{}l@{}}
$\llbracket 1;\mathsf{N} \rrbracket$ & $\rightarrow\{1,2,\hdots,N\}$ \\
$s^{n}$ & $\rightarrow$ the $n^{th}$ term of a sequence $s=(s^{1},s^{2},\hdots)$ \\
$v_{i}$ & $\rightarrow$ the $i^{th}$ component of an array $v=(v_{1},v_{2},\hdots)$\\
$f^{k}$ & $\rightarrow$ $k^{th}$ composition of a function $f$ \\
& $\begin{array}{r@{\;}l}\ f^{k}=\underbrace{f\circ ...\circ f} \\ \ k\ \text{times}\end{array}$\\
$\emph{strategy}$ & $\rightarrow$ a sequence which elements belong in $%
\llbracket 1;\mathsf{N} \rrbracket $ \\
$\mathbb{S}$ & $\rightarrow$ the set of all strategies \\
$\oplus$ & $\rightarrow$ bitwise exclusive or \\
$+$ & $\rightarrow$ the integer addition \\
$\ll \text{and} \gg$ & $\rightarrow$ the usual shift operators \\
\end{tabular}
\subsection{Chaotic iterations}
\label{subsection:Chaotic iterations}
\begin{definition}
The set $\mathds{B}$ denoting $\{0,1\}$, let $f:\mathds{B}^{\mathsf{N}%
}\longrightarrow \mathds{B}^{\mathsf{N}}$ be an ``iteration'' function and $S\in \mathbb{S}
$ be a chaotic strategy. Then, the so-called \emph{chaotic iterations} are defined by
\begin{equation}
\begin{array}{l}
x^0\in \mathds{B}^{\mathsf{N}}, \\
\forall n\in \mathds{N}^{\ast },\forall i\in \llbracket1;\mathsf{N}\rrbracket%
,x_i^n=\left\{
\begin{array}{l}
x_i^{n-1} ~~~~~\text{if}~S^n\neq i \\
f(x^{n-1})_{S^n} ~\text{if}~S^n=i.\end{array} \right. \end{array}
\end{equation}
\end{definition}
In other words, at the $n^{th}$ iteration, only the $S^{n}-$th cell is
\textquotedblleft iterated\textquotedblright . Note that in a more general
formulation, $S^n$ can be a subset of components and $f(x^{n-1})_{S^{n}}$ can
be replaced by $f(x^{k})_{S^{n}}$ (where $k < n$), thus describing for
example, delays due to transmissions (see \emph{e.g.}~\cite{Bahi2000}). For the
general definition of such chaotic iterations, see, e.g.~\cite{Robert1986}.\newline
Chaotic iterations generate a set of vectors (boolean vector in this paper),
they are defined by an initial state $x^{0}$, an iteration function $f$ and a
chaotic strategy $S$.
\subsection{XORshift}
XORshift is a category of very fast PRNGs designed by George Marsaglia~\cite{Marsaglia2003}. It repeatedly uses the exclusive or (XOR) on a number, with a bit shifted copy of itself by $a$ positions either to the right or to the left, where $ 0 < a < w$ and $w = 32$ or 64. The initial state of a XORshift generator is a given vector of bits. At each step, the next state is obtained by applying a given number of XORshift operations as defined in Table~\ref{XORshift}. This algorithm has a period of $2^{32}-1=4.29\times10^9$.

\begin{table}
\label{XORshift}
\label{Chaotic iteration}
\centering
\begin{tabular}{|l|}
\hline
~\textbf{Input}: $x$ (a 32-bit word) \\
\hline
~\textbf{Output}: $r$ (a 32-bit word)\\
\hline
~$x\leftarrow{x\oplus{(x\ll13)}}$;\\
~$x\leftarrow{x\oplus{(x\gg17)}}$;	\\
~$x\leftarrow{x\oplus{(x\ll5)}}$;	\\
~$r\leftarrow{x}$;			\\
~\textbf{return} $r$;			\\

\hline
~\textbf{An arbitrary round of XORshift}~ \\
\hline

\end{tabular}
\caption{XORshift and CI algorithms}
\label{CI}\label{XORshift}
\end{table}

\subsection{Input sequences}
In \cite{wang2009}, we have used two logistic maps~\cite{ulam1947} as input sequences to define a novel PRNG (called CI PRNG) based on chaotic iterations. We have mathematically proven that it behaves chaotically, as defined by Devaney. In addition, this generator can successfully pass the NIST tests suite. However, chaotic systems like logistic maps work in the real numbers domain, and therefore a transformation from real numbers into integers is needed. This process leads to a degradation of the chaotic behavior of the generator and a lot of time wasted during computations~\cite{ALI2007}. Our purpose is then to improve the speed of this former generator and grant its chaotic properties by using a faster PRNG, namely XORshift. Moreover, we will show in Section~\ref{Subsec:DieHARD} that this new generator can pass the famous DieHARD battery of tests.
\section{The generation of CI pseudo-random sequence}
\label{The generation of pseudo-random sequence}
The design of the PRNG based on discrete chaotic iterations is proposed in this section, while its performance is evaluated in the next one.
\subsection{Chaotic iterations as PRNG}
The novel generator is designed by the following process.
Let $\mathsf{N} \in \mathds{N}^*, \mathsf{N} \geqslant 2$. Some chaotic iterations are done, which generate a sequence $\left(x^n\right)_{n\in\mathds{N}} \in \left(\mathds{B}^\mathsf{N}\right)^\mathds{N}$ of boolean vectors: the successive states of the iterated system. Some of those vectors are chaotically extracted and their components constitute our pseudo-random bit flow.
Chaotic iterations are realized as follows: initial state\linebreak $x^0 \in \mathds{B}^\mathsf{N}$ is a boolean vector taken as a seed and chaotic strategy $\left(S^n\right)_{n\in\mathds{N}}\in \llbracket 1, \mathsf{N} \rrbracket^\mathds{N}$ is constructed with XORshift. Lastly, iterate function $f$ is the vectorial boolean negation
$$f_0:(x_1,...,x_\mathsf{N}) \in \mathds{B}^\mathsf{N} \longmapsto (\overline{x_1},...,\overline{x_\mathsf{N}}) \in \mathds{B}^\mathsf{N}.$$
To sum up, at each iteration, only $S^i$-th component of state $x^n$ is updated as follows
\begin{equation}
x_i^n = \left\{\begin{array}{ll}x_i^{n-1} & \text{if } i \neq S^i, \\ \\ \overline{x_i^{n-1}} & \text{if } i = S^i. \\\end{array}\right.
\end{equation}
Finally, let $\mathcal{M}$ be a finite subset of $\mathds{N}^*$. Some $x^n$ are selected by a sequence $m^n$ as the pseudo-random bit sequence of our generator. The sequence $(m^n)_{n \in \mathds{N}} \in \mathcal{M}^\mathds{N}$ is computed with XORshift. So, the generator returns the following values:
\begin{itemize}
\item the components of $x^{m^0}$,
\item following by the components of $x^{m^0+m^1}$,
\item following by the components of $x^{m^0+m^1+m^2}$,
\item \emph{etc.}
\end{itemize}
In other words, the generator returns the following bits:\newline
\begin{small}
$$x_1^{m_0}x_2^{m_0}x_3^{m_0}\hdots x_\mathsf{N}^{m_0}x_1^{m_0+m_1}x_2^{m_0+m_1}\hdots x_\mathsf{N}^{m_0+m_1} x_1^{m_0+m_1+m_2}x_2^{m_0+m_1+m_2}\hdots$$
\end{small}
and its $k^{th}$ bit is equal to $$\displaystyle{x_{k+1 \text{ (mod }\mathsf{N}\text{)}}^{\sum_{i=0}^{\lfloor k/\mathsf{N} \rfloor}m_i}}.$$
The basic design procedure of the novel generator is summed up in Table~\ref{CI}.
The internal state is $x$, the output array is $r$; $a$ and $b$ are those computed by the two XORshift generators. Lastly, $c$ and $\mathsf{N}$ are constants and $\mathcal{M}=\{\text{c,c+1}\}$ ($c\geqslant 3\mathsf{N}$ is recommended).
\begin{table}
\centering
\begin{tabular}{|l|}
\hline
~\textbf{Input}: the internal state $x$ \\
~~~~~~~~~~~($x$ is an array of $\mathsf{N}$ 1-bit words) \\
\hline
~\textbf{Output}: an array $r$ of $\mathsf{N}$ 1-bit words\\
\hline
~$a\leftarrow{XORshift1()}$;\\
~$m\leftarrow{a~mod~2+c}$\\
~\textbf{for} $i=0,\dots,m$ \textbf{do}\\
~~~~~~$b\leftarrow{XORshift2()}$;\\
~~~~~~$S\leftarrow{b~mod~\mathsf{N}}$;\\
~~~~~~$x_S\leftarrow{ \overline{x_S}}$;\\
~\textbf{end for}\\
~$r\leftarrow{x}$;\\
~\textbf{return} $r$;\\
\hline
~\textbf{An arbitrary round of CI generator}~\\
\hline

\end{tabular}
\caption{CI algorithms}
\label{CI}
\end{table}
\subsection{Example}
In this example, $\mathsf{N} = 5$ and $\mathcal{M} = \{\text{4,5}\}$ are adopted for easy understanding.
The initial state of the system $x^0$ can be seeded by the decimal part of the current time. For example, the current time in seconds since the Epoch is 1237632934.484084, so $t = 484084$. $x^0 = t \text{ (mod 32)}$ in binary digits, then $x^0 = (1, 0, 1, 0, 0)$. $m$ and $S$ can now be computed from two XORshift PRNGs:

$m$ = 4, 5, 4, 4, 4, 4, 5, 5, 5, 5, 4, 5, 4,...

$S$ = 2, 4, 2, 2, 5, 1, 1, 5, 5, 3, 2, 3, 3,...

Chaotic iterations are made with initial state $x^0$, vectorial logical negation $f_0$ and strategy $S$. The result is presented in Table \ref{table application example}. Let us recall that sequence $m$ gives the states $x^n$ to return: $x^4, x^{4+5}, x^{4+5+4}, \hdots$\newline
\begin{tiny}
\begin{table*}[!t]
\renewcommand{\arraystretch}{1.3}
\caption{Application example}
\label{table application example}
\centering
\begin{tabular}{c|ccccc|cccccc|cccccc}
\hline\hline
$m:$ & & & 4 & & & & & 5 & & & & & & 4 & & & \\ \hline
$S$ & 2 & 4 & 2 & 2 & & 5 & 1 & 1 & 5 & 5 & & 3 & 2 & 3 & 3 & & \\ \hline
$x^{0}$ & & & & & $x^{4}$ & & & & & & $x^{9}$ & & & & & $x^{13}$ & \\
1 & & & & &
1 & & $\xrightarrow{1} 0$ & $\xrightarrow{1} 1$ & & &
1 & & & & &
1 & \\
0 & $\xrightarrow{2} 1$ & & $\xrightarrow{2} 0$ & $\xrightarrow{2} 1$ &
1 & & & & & &
1 & & $\xrightarrow{2} 0$ & & & 0 &\\
1 & & & & &
1 & & & & & &
1 & $\xrightarrow{3} 0$ & & $\xrightarrow{3} 1$ & $\xrightarrow{3} 0$ &
0 &\\
0 & & $\xrightarrow{4} 1$ & & &
1 & & & & & &
1 & & & & &
1 &\\
0 & & & & &
0 & $\xrightarrow{5} 1$ & & & $\xrightarrow{5} 0$ & $\xrightarrow{5} 1$ &
1 & & & & &
1 &\\
\hline\hline
\end{tabular}\\
\vspace{0.5cm}
Output: $x_1^{0}x_2^{0}x_3^{0}x_4^{0}x_5^{0}x_1^{4}x_2^{4}x_3^{4}x_4^{4}x_5^{4}x_1^{9}x_2^{9}x_3^{9}x_4^{9}$
$x_5^{9}x_1^{13}x_2^{13}x_3^{13}x_4^{13}x_5^{13}... = 10100111101111110011...$
\end{table*}
\end{tiny}
So, in this example, the output of the generator is: 10100111101111110011...
\subsection{Chaotic properties}
Despite a large number of papers published in the field of chaos-based PRNGs, the impact that this research has made on conventional information security is rather marginal. This is due to the following reasons: almost all chaotic algorithms are based on dynamical systems defined on the set of real numbers. So these generators are usually slow, require considerably more storage space and lose some of their chaotic properties during computations. These major problems restrict their use in security fields as cryptography~\cite{Kocarev2001}.\newline
The PRNG proposed in this paper does not inherit its chaotic properties from a real chaotic map, but from chaotic iterations defined in Section \ref{subsection:Chaotic iterations}. It has been proven in~\cite{guyeux09} that CIs behave as chaos, as it is defined by Devaney: they are regular, transitive and sensitive to initial conditions. This most famous definition of chaotic behavior for a dynamical system implies various desired properties in information security, such as: unpredictability, mixture, sensitivity, and uniform repartition. The principal interest of CIs is that they can be used without real numbers. Indeed, the sequence inputed in chaotic iterations constitutes a coordinate of its initial state, and the chaotic behavior of a dynamical system does not depend on this initial state. So if we take integer sequences as input instead, then CIs become faster while preserving their chaotic properties. This allows the conception of a new generation of fast and chaotic PRNGs
\section{Testing a generator}
\label{Testing a generator}
Here, the empirical tests have been carried out, making use of the DieHARD statistical test suite. In this section we will briefly review the approach taken together with key results and conclusions. It is not our intention to document these tests in detail in the present section, since it has been done several times in many other papers~\cite{Marsaglia1996},\cite{Tan2009},\cite{Turan2008}.
\subsection{DieHARD battery of tests}
DieHARD battery of tests has been a stringent standard for evaluating PRNGs for over a decade. Passing this battery is considered as a good rule of thumb to validate a PRNG. DieHARD battery consists of 18 different independent statistical tests. This collection of tests is based on assessing the randomness of bits comprising 32-bit integers obtained from a random number generator. Each test requires $2^{23}$ 32-bit integers in order to run the full set of tests.\newline
Most of the tests in DieHARD return a $p-$value, which should be uniform on $[0,1)$ if the input file contains truly independent random bits. Those $p-$values are obtained by $p=F(X)$, where $F$ is the assumed distribution of the sample random variable $X$\textendash often supposed as normal distribution. But that assumed $F$ is just an asymptotic approximation, for which the fit will be worst in the tails. Thus occasional $p-$values near 0 or 1, such as 0.0012 or 0.9983, are not surprising. An individual test is considered to be a failure if its $p-$value approaches 1 more closely, for example $p>0.9999$.
\subsection{Analysis}
\label{Subsec:DieHARD}
Table~\ref{Results of DieHARD battery of tests} gives the results derived from applying the DieHARD battery of tests to the RNGs considered in this work. As it can be observed, the results of the individual tests Count the ones 1, Binary Rank $31 \times 31$ and Binary Rank $32 \times 32$ show that in the random numbers obtained with the XORshift generator only the least significant bits seem to be independent. This explains the poor behavior of this RNG in the aforementioned basic tests that evaluate the independence of real numbers. But the generator based on discrete chaotic iterations can pass all the DieHARD battery of tests. 
This proves that the
security of the given generator has been improved by CIs.
\begin{tiny}
\begin{table}[!t]
\renewcommand{\arraystretch}{1.3}
\caption{Results of DieHARD battery of tests}
\label{Results of DieHARD battery of tests}
\centering
\begin{tabular}{llcc} \hline
\textbf{No.} &\textbf{Test name} &\multicolumn{2}{c}{\textbf{Generators}} \\ \cmidrule(r){3-4}
& & XORshift & PRNG (Chaotic iterations)\\ \hline
1 & Overlapping Sum &Pass &Pass \\
2 & Runs Up 1 &Pass & Pass \\
&Runs Down 1 &Pass &Pass \\
&Runs Up 2 & Pass &Pass \\
&Runs Down 2 &Pass & Pass \\
3 & 3D Spheres &Pass &Pass \\
4 & Parking Lot &Pass &Pass \\
5 & Birthday Spacing &Pass &Pass \\
6 & Count the ones 1 &Fail &Pass \\
7 &Binary Rank $6 \times 8$ &Pass & Pass \\
8 &Binary Rank $31 \times 31$ &Fail &Pass \\
9 &Binary Rank $32 \times 32$ &Fail &Pass \\
10 &Count the ones 2 &Pass &Pass \\
11 &Bit Stream &Pass &Pass \\
12 &Craps Wins &Pass &Pass \\
&Throws &Pass &Pass \\
13 &Minimum Distance &Pass &Pass \\
14 &Overlapping Perm. &Pass &Pass \\
15 &Squeeze &Pass &Pass \\
16 &OPSO &Pass &Pass \\
17 &OQSO &Pass &Pass \\
18 &DNA &Pass &Pass \\
&Number of tests passed &15 &18 \\\hline
\end{tabular}
\end{table}
\end{tiny}
%
\section{Application example in digital watermarking}
\label{An application example of the proposed PRNG}
Information hiding has recently become a major information security technology, especially with the increasing importance and widespread distribution of digital media through the Internet \cite{Wu2007bis}. It includes several techniques like digital watermarking. The aim of digital watermarking is to embed a piece of information into digital documents, such as pictures or movies. This is for a large panel of reasons, such as: copyright protection, control utilization, data description, content authentication, and data integrity. For these reasons, many different watermarking schemes have been proposed in recent years. Digital watermarking must have essential characteristics, including: security, imperceptibility, and robustness. 
Chaotic methods have been proposed to encrypt the watermark before embedding it in the carrier image for these security reasons. 
In this paper, a new watermarking algorithm is given. It is based on the chaotic PRNG presented above.
\subsection{Most and least significant coefficients}
Let us first introduce the definitions of most and least significant coefficients.
\begin{definition}
For a given image, the most significant coefficients (in short MSCs), are
coefficients that allow the description of the relevant part of the image,
\emph{i.e.} its most rich part (in terms of embedding information), through
a sequence of bits.
\end{definition}
For example, in a spatial description of a grayscale image, a definition of
MSCs can be the sequence constituted by the first three bits of each pixel as
shown in Figure 1(a).
In a discrete cosine frequency domain description, each $8\times 8$ block of
the carrier image is mapped to a list of 64 coefficients. The energy of the
image is contained in the first of them. After binary conversion, the
first fourth coefficients of all these blocks can constitute a possible
sequence of MSCs.
\begin{definition}
By least significant coefficients (LSCs), we mean a translation of some
insignificant parts of a medium in a sequence of bits (insignificant can be
understand as: ``which can be altered without sensitive damages'').
\end{definition}
These LSCs can be for example, the last three bits of the gray level of each
pixel, in the case of a spatial domain watermarking of a grayscale image, as in
Figure 1(b).

\begin{center}
\begin{tabular}{cc}
\includegraphics[scale=0.3]{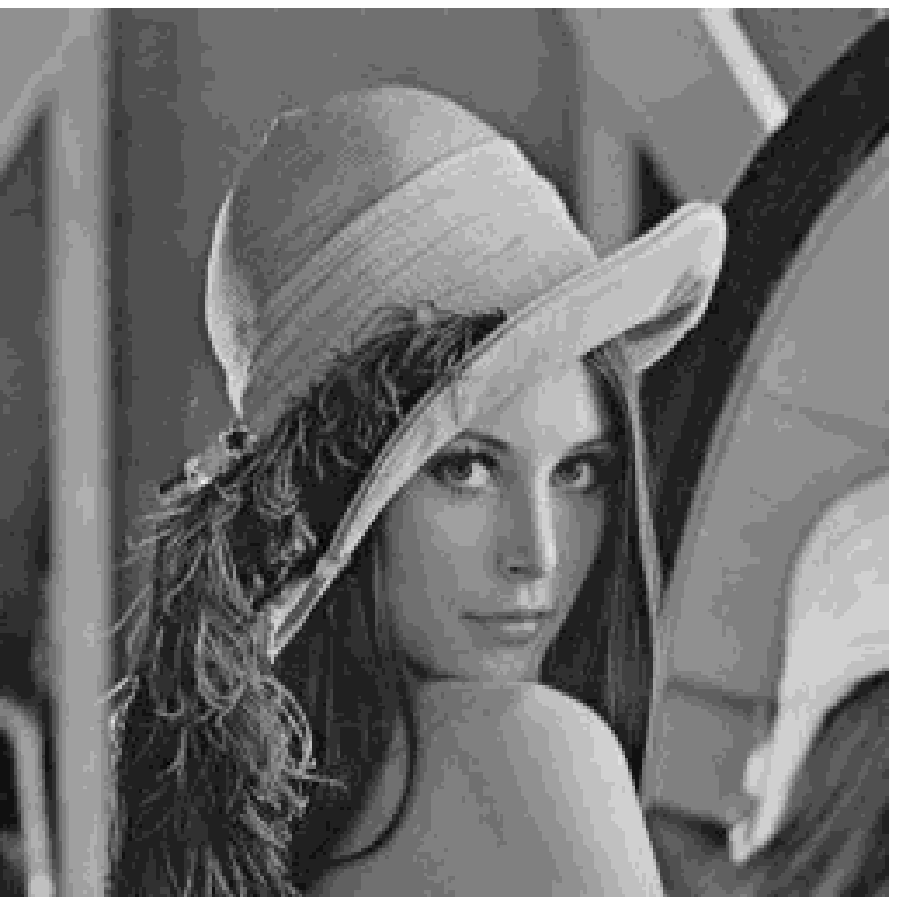} &~~~~~~~~~~~~~~~~ \includegraphics[scale=0.3]{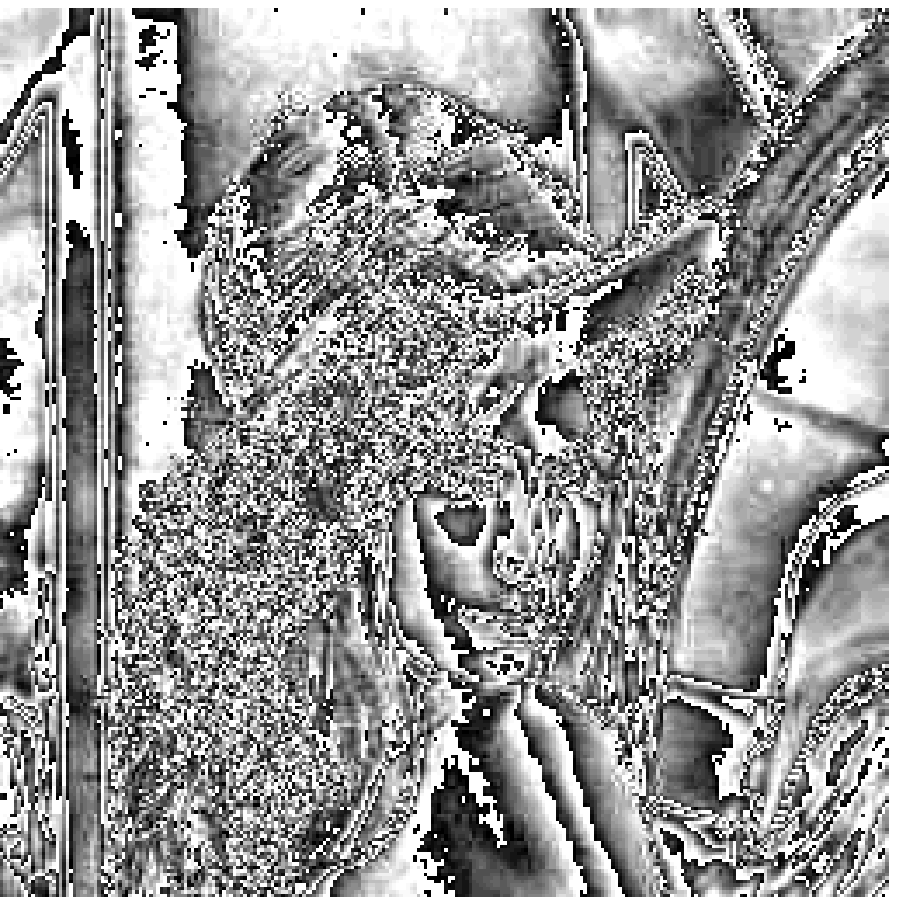}\\
(a) MSCs of Lena &~~~~~~~~~~~~~~~~(b) LSCs of Lena   \\ 
\end{tabular}
Figure 1. Spatial MSCs and LSCs of Lena.
\end{center}

Discrete cosine, Fourier, and wavelet transform can be used to define LSCs
and MSCs, in the case of frequency domain watermarking, among other possible choices. Moreover, these definitions are not limited to image media, but
can easily be extended to the audio and video media as well. \newline
LSCs are used during the embedding stage: some of the least significant
coefficients of the carrier image will be chaotically chosen and replaced by
the bits of the mixed watermark. With a large number of LSCs,
the watermark can be inserted more than once and thus the embedding will
be more secure and robust, but also more detectable.\newline
The MSCs are only useful in the case of authentication: encryption and embedding
stages depend on them. Hence, a coefficient should not be defined
at the same time, as a MSC and a LSC; the last can be altered, while the
first is needed to extract the watermark.
\subsection{Stages of the algorithm}
Our watermarking scheme consists of two stages: (1) mixture of the watermark and (2) its embedding.
\subsubsection{Watermark mixture.}
Firstly, for safety reasons, the watermark can be mixed before its embedding into the image. A common way to achieve this stage is to use the bitwise exclusive or (XOR), for example, between the watermark and the above PRNG. In this paper, we will use another mixture scheme based on chaotic iterations. Its chaotic strategy, defined with our PRNG, will be highly sensitive to the MSCs, in the case of an authenticated watermark, as stated in ~\cite{guyeux09}. 

\subsubsection{Watermark embedding.}
Some LSCs will be substituted by all bits of the possibly mixed watermark. To choose the sequence of LSCs to be altered, a number of integers, less than or equal to the number $\mathsf{N}$ of LSCs corresponding to a chaotic sequence $\left( U^{k}\right) _{k}$, is generated from the chaotic strategy used in the mixture stage. Thus, the $U^{k}$-th least significant coefficient of the carrier image is substituted by the $k^{th}$ bit of the possibly mixed watermark. In the case of authentication, such a procedure leads to a choice of the LSCs which are highly dependent on the MSCs.
For the detail of this stage see Section \ref{Watermark embedding}.
\subsubsection{Extraction.}
The chaotic strategy can be regenerated, even in the case of an authenticated watermarking because the MSCs have not been changed during the stage of embedding the watermark. Thus, the few altered LSCs can be found, the mixed watermark can then be rebuilt, and the original watermark can be obtained.
If the watermarked image is attacked, then the MSCs will change. Consequently, in the case of authentication and due to the high sensitivity of the embedding sequence, the LSCs designed to receive the watermark will be completely different. Hence, the result of the recovery will have no similarity with the original watermark: authentication is reached.
\section{Evaluation of robustness}
\label{Robustness evaluations}
In this section, a complete application example of the above chaotic
watermarking method is given and its robustness to some attacks is studied.
This case study enables us to precise the details of the algorithm and evaluate it.
\subsection{Stages and details}
\subsubsection{Images description.}
\label{Images description} Carrier image is Lena, a
256 grayscale image of size $256 \times 256$. The watermark is the $64\times 64$ pixels binary image depicted in 
Figure 2(a). The embedding domain will be the spatial domain. The
selected MSCs are the four most significant bits of each pixel and the LSCs
are the three last bits (a given pixel will at most be modified of four
levels of gray by an iteration). Before its
embedment, the watermark is mixed with chaotic iterations. The system to
iterate, chaotic strategy $S^n$ and iterate function are defined below.
\begin{center}
\begin{tabular}{cc}
\multicolumn{2}{c}{\includegraphics[scale=1]{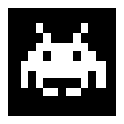}}  \\
\multicolumn{2}{c}{(a)Watermark} \\
\includegraphics[scale=0.18]{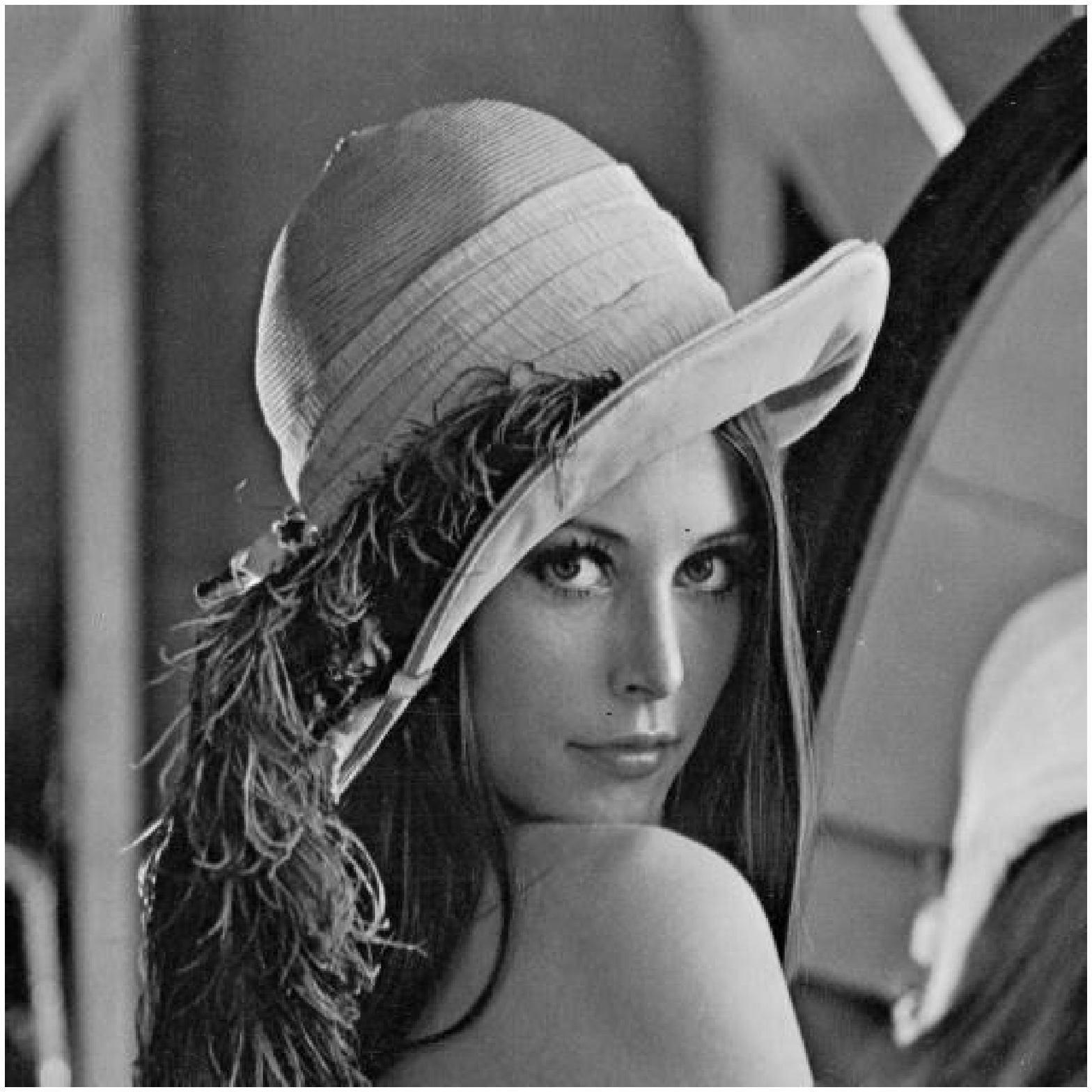}&\includegraphics[scale=0.36]{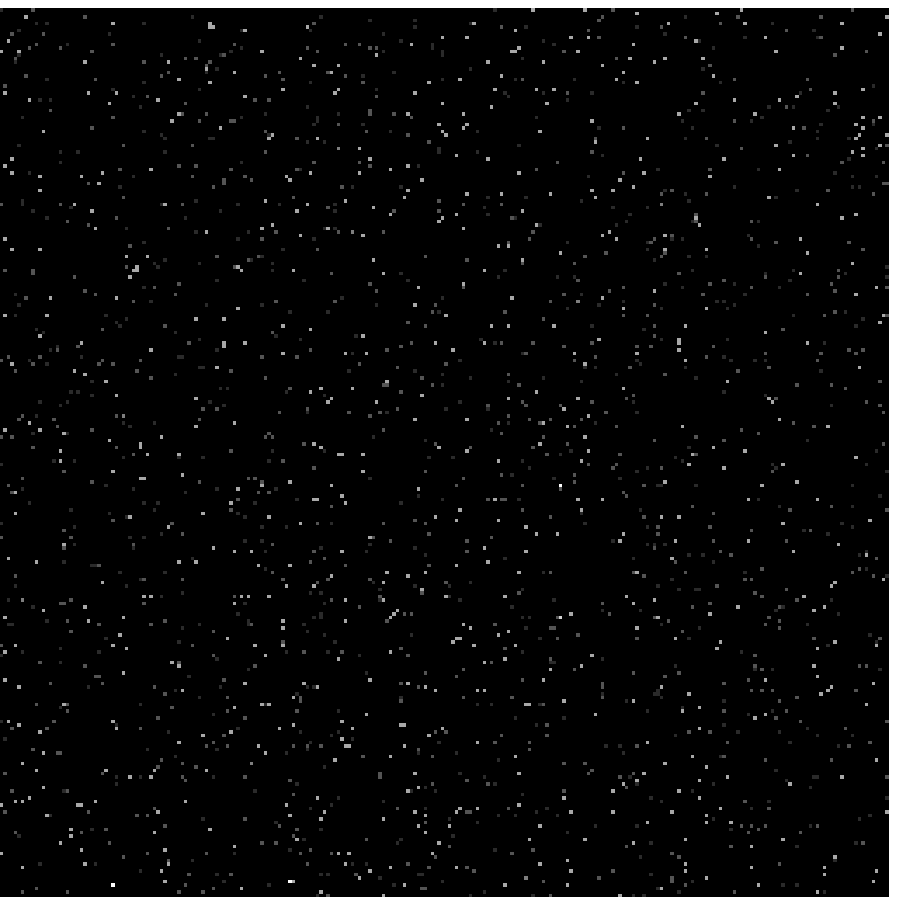} \\
(b)Watermarked Lena  &(c)Differences with original.\\ 
\end{tabular}
Figure 2. Watermarked Lena and differences
\label{fig:LenaWatermark}
\end{center}

\subsubsection{Embedding of the watermark.}
\label{Watermark embedding}
To embed the watermark, the sequence $(U^{k})_{k\in \mathds{N}}$ of altered
bits taken from the $\mathsf{M}$ LSCs must be defined. To do so, the strategy $%
(S^{k})_{k\in \mathds{N}}$ of the encryption stage is used as follows
\begin{equation}
\left\{
\begin{array}{lll}
U^{0} & = & S^{0} \\
U^{n+1} & = & S^{n+1}+2\times U^{n}+n ~(mod ~ \mathsf{M})
\end{array}%
\right.
\end{equation}
to obtain the result depicted in Figure 2(b). The map $%
\theta \mapsto 2\theta $ of the torus, which is a famous example of
topological Devaney's chaos \cite{Dev89}, has been chosen to make $%
(U^{k})_{k\in \mathds{N}}$ highly sensitive to the chaotic strategy $(S^{k})_{k\in \mathds{N}}$. As a
consequence, $(U^{k})_{k\in \mathds{N}}$ is highly sensitive to the alteration
of the MSCs. In case of authentication, any significant modification of the
watermarked image will lead to a completely different extracted watermark.
\subsection{Simulation results}
To prove the efficiency and the robustness of the proposed
algorithm, some attacks are applied to our chaotically watermarked image. For
each attack, a similarity percentage with the original watermark is computed. This
percentage is the number of equal bits between the original and the
extracted watermark, shown as a percentage. A result less
than or equal to $50\%$ implies that the image has probably not been
watermarked.
\subsubsection{Cropping attack.}
In this kind of attack, a watermarked image is cropped. In this case, the
results in Table~\ref{Cropping attacks} have been obtained.

In Figure 3, the decrypted watermarks are shown
after a crop of 50 pixels and after a crop of 10 pixels, in the
authentication case.
\begin{center}
\begin{tabular}{cc}
\includegraphics[scale=0.8]{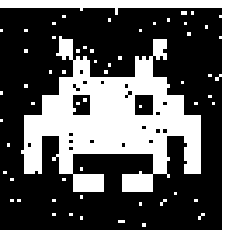} & \includegraphics[scale=0.8]{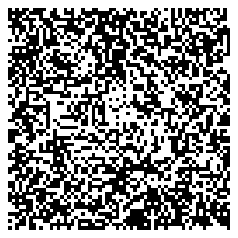}\\ 
(a)Unauthentication ($10\times 10$) &(b)Authentication($10\times 10$)\\
\multicolumn{2}{c}{\includegraphics[scale=0.8]{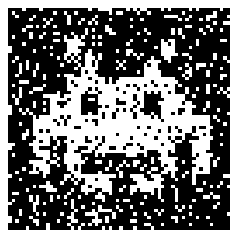}} \\

\multicolumn{2}{c}{(c)Unauthentication   ($50\times 50$) }\\

\end{tabular}
Figure 3. Extracted watermark after a cropping attack.
\end{center}

By analyzing the similarity percentage between the original and the
extracted watermark, we can conclude that in the case of unauthentication, the
watermark still remains after a cropping attack. The desired robustness is
reached. It can be noticed that cropping sizes and percentages are rather
proportional. In the case of authentication, even a small change of the carrier
image (a crop by $10\times 10$ pixels) leads to a really different
extracted watermark. In this case, any attempt to alter the carrier image
will be signaled, thus the image is well authenticated.
\subsubsection{Rotation attack.}
Let $r_{\theta }$ be the rotation of angle $\theta $ around the center $%
(128, 128)$ of the carrier image. So, the transformation $r_{-\theta }\circ
r_{\theta }$ is applied to the watermarked image. The results in Table~\ref{Rotation attacks}
have been obtained.
The same conclusion as above can be declaimed.
\subsubsection{JPEG compression.}
A JPEG compression is applied to the watermarked image, depending on a
compression level. This attack leads to a change of
the representation domain (from spatial to DCT domain). In this case, the
results in Table~\ref{JPEG compression attacks} have been obtained, illustrating a good authentication through JPEG attack.
As for the unauthentication case, the watermark still remains after a compression level equal to 10. This is a good result if we take into account the fact that we use spatial embedding.
\subsubsection{Gaussian noise.}
A watermarked image can be also attacked by the addition of a Gaussian noise,
depending on a standard deviation. In this case, the results in Table \ref{Gaussian noise attacks} are
obtained.
\begin{table}
\caption{Attacks}
\label{Cropping attacks}
\label{Rotation attacks}
\label{JPEG compression attacks}
\label{Gaussian noise attacks}
\begin{center}
{\footnotesize
\begin{tabular}{c c c c c}
\toprule
Attacks&\multicolumn{2}{c}{UNAUTHENTICATION} & \multicolumn{2}{c}{AUTHENTICATION}
\\ \midrule 
\multirow{5}*{\rotatebox{90}{Cropping}} &Size (pixels) & Similarity & Size (pixels) & Similarity \\ \cmidrule(r){2-5}
		&10 & 99.48\% & 10 & 49.68\% \\
		&50 & 97.63\% & 50 & 54.54\% \\
		&100 & 91.31\% & 100 & 52.24\% \\
		&200 & 68.56\% & 200 & 51.87\% \\  \midrule 
\multirow{5}*{\rotatebox{90}{Rotation } }		&Angle (degree) & Similarity & Angle (degree) & Similarity \\ \cmidrule(r){2-5}
		&2 & 97.41\% & 2 & 70.01\% \\
	&5 & 94.67\% & 5 & 59.47\% \\
		&10 & 91.30\% & 10 & 54.51\% \\
		&25 & 80.85\% & 25 & 50.21\% \\ \midrule 
\multirow{5}*{\rotatebox{90}{JPEG compression}}	&&&&\\
	&Compression & Similarity & Compression & Similarity \\ \cmidrule(r){2-5}
		&2 & 82.95\% & 2 & 54.39\% \\
&5 & 65.23\% & 5 & 53.46\% \\
		&10 & 60.22\% & 10 & 50.14\% \\
		&20 & 53.17\% & 20 & 48.80\% \\ 
&&&&\\
\midrule 
\multirow{5}*{\rotatebox{90}{Gaussian noise}}	\\

		&Standard dev. & Similarity & Standard dev. & Similarity \\ \cmidrule(r){2-5}
		&1 & 74.26\% & 1 & 52.05\% \\
&2 & 63.33\% & 2 & 50.95\% \\
		&3 & 57.44\% & 3 & 49.65\% \\ \bottomrule
\end{tabular}
}\\[0pt]
\end{center}
\end{table}

\section{Conclusions and future work}
\label{Conclusions and Future Work}
In this paper, the PRNG proposed in \cite{wang2009} is improved, by using the famous XORshift generator. By combining these components with chaotic iterations, we define a faster generator with chaotic properties. In addition to achieving the NIST tests suite, this new generator successfully passes all the stringent DieHARD battery of tests. The randomness and disorder generated by this algorithm has been evaluated. It offers a sufficient level of security for a whole range of computer usages. An application example in the field of data hiding is given and its robustness through attacks is studied. In future work, the speed of our generator will be improved again, the comparison of different chaotic strategies will be explored, and other iteration functions will be studied. Finally, new applications in computer science security field will be proposed.
\bibliographystyle{plain}
\bibliography{Generating_good_chaotic_random_numbers.bib}
\end{document}